\documentclass[aps,prb,twocolumn,showpacs,preprintnumbers,amsmath,amssymb]{revtex4}

\usepackage{graphicx}
\usepackage{dcolumn}
\usepackage{bm}

\begin{document}


\title{Detrapping and retrapping of free carriers in nominally pure single crystal GaP, GaAs and 4H-SiC semiconductors under light illumination at cryogenic temperatures}

\author{David Mouneyrac,$^{1,2}$  John G. Hartnett,$^1$  Jean-Michel Le Floch,$^1$ Michael E. Tobar,$^1$ Dominique Cros,$^2$ Jerzy Krupka$^3$}
 \email{john@physics.uwa.edu.au}
\affiliation{
$^1$School of Physics, University of Western Australia 35 Stirling Hwy, Crawley 6009 WA Australia\\
$^2$Xlim, UMR CNRS 6172, 123 av. Albert Thomas, 87060 Limoges Cedex - France\\
$^3$Institute of Microelectronics and Optoelectronics Department of Electronics, Warsaw University of Technology, Warsaw, Poland\\
}

\date{\today}

\begin{abstract}
We report on extremely sensitive measurements of changes in the microwave properties of high purity non-intentionally-doped single-crystal semiconductor samples of gallium phosphide, gallium arsenide and 4H-silicon carbide when illuminated with light of different wavelengths at cryogenic temperatures.  Whispering gallery modes were excited in the semiconductors whilst they were cooled on the coldfinger of a single-stage cryocooler and their frequencies and Q-factors measured under light and dark conditions.  With these materials, the whispering gallery mode technique is able to resolve changes of a few parts per million in the permittivity and the microwave losses as compared with those measured in darkness. A phenomenological model is proposed to explain the observed changes, which result not from direct valence to conduction band transitions but from detrapping and retrapping of carriers from impurity/defect sites with ionization energies that lay in the semiconductor band gap. Detrapping and retrapping relaxation times have been evaluated from comparison with measured data. 
\end{abstract}

\maketitle

\section{Introduction}
Since 1977, many experiments have been performed in order to study the effects of light on semiconductors such as Zn$_{0.02}$Cd$_{0.98}$Te, \cite{Dissanayake2, Dissanayake3} AlAs/GaAs, \cite{Jeanjean} XAs:Si, \cite{Brunthaler} AlGaN/GaN, \cite{Dang} Al$_x$Ga$_{1-x}$C$_{60}$, \cite{Hamed} gallium phosphide (GaP) and gallium arsenide (GaAs). \cite{Hartnett1, Hartnett2}  Those experiments showed changes in the microwave properties of the semiconductors under light illumination and persistence of photoconductivity after the illumination was terminated. 

Different models have been developed to explain these effects \cite{Lang, Nelson, Jiang, Dissanayake1} but the exact mechanism still remains a controversy. Previously we presented the results of measurements of gallium phosphide and gallium arsenide under illumination using the  very sensitive ``whispering gallery mode'' method. \cite{Hartnett1, Hartnett2}  The whispering gallery mode technique used on these samples is able to resolve changes in microwave permittivity and losses of the order of a few parts per million. That allowed us to precisely determined the time constants for the trapping and de-trapping rates in gallium phosphide and gallium arsenide under white light illumination.  

In that previous work\cite{Hartnett1} double time constants were observed. Since then, more measurements have been made to try to understand the phenomenon.  It was discovered that one of the time constants, of the two observed, was related to a warming up of the halogen lamp, not a heating effect on the resonator, but an increase intensity of the source as it started up. That introduced a spurious effect to the evaluation of the relaxation times under light illumination in those materials. We have now made rigorous measurements, wherein we have eliminated this effect by allowing the halogen lamp to come to full intensity before connecting the optical fiber to the resonant cavity. And in the case of the lasers used in this work a shutter was used to interrupt the beam.

In this paper we extend the previous work to include also illumination with three different monochromatic sources, and over a larger cryogenic temperature range.  And we develop a new hypothesis to explain the role of  impurity and defect sites with ionization energies in the forbidden band gap of these semiconductors, GaP, GaAs \cite{Queisser, Krupka} and 4H-silicon carbide (4H-SiC)\cite{Hartnett3,Persson} at microwave frequencies and cryogenic temperatures. We apply a simple mathematical model that precisely describes detrapping and retrapping of carriers as a function of time from switching the light source on or off. This phenomenological model is consistent with all measurements at different wavelengths of incident light and at different temperatures.
\begin{figure}
\includegraphics[width = 3 in]{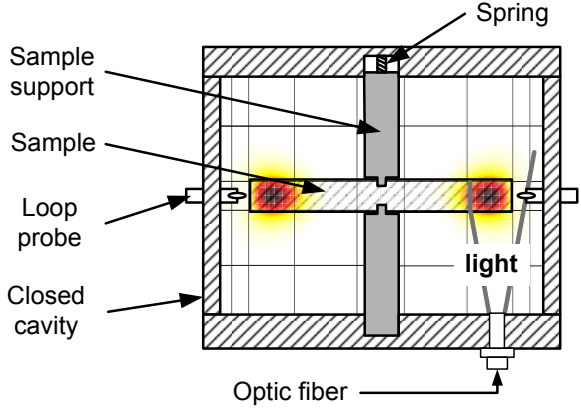}
\caption{\label{fig:fig1} Semiconductor sample located in a closed cavity and excited by two magnetic loop probes connected via coaxial cables to a vector network analyzer in the room temperature environment.}
\end{figure}
\begin{figure}
\includegraphics[width = 3.5 in]{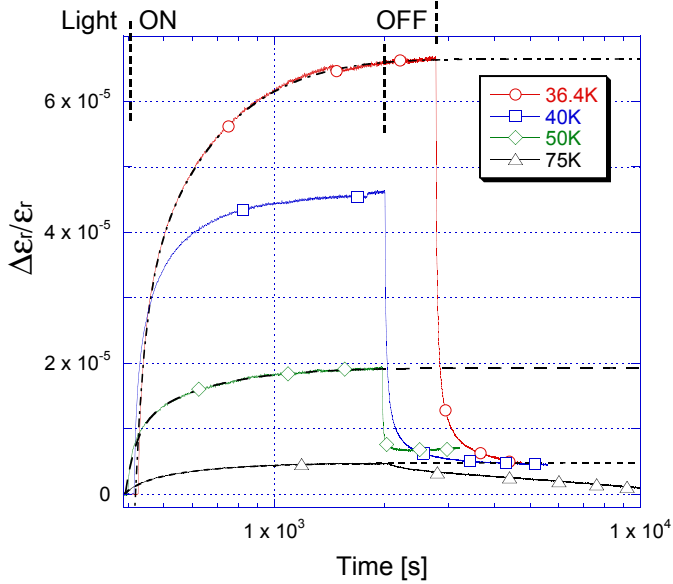}
\caption{\label{fig:fig2} (Color online) Time evolution of the relative permittivity ($\varepsilon_r$) of GaP illuminated with a 10 mW white light source at different temperatures. The (black) dashed and dot-dashed lines are modeled fits. See text for details.} 
\end{figure}

\section{Method}
High-purity non-doped single-crystal semiconductor cylinders of GaP (48.1 mm diameter and 5 mm height), GaAs (25.4 mm diameter and 6.3 mm height) and 4H-SiC (11 mm diameter and 2.62 mm height) were placed in silver-plated copper cavities (see Fig. 1; individual cavity dimensions were chosen to suit the semiconductor sample under examination \cite{Hartnett1, Hartnett2, Hartnett3}) located in a vacuum chamber and cooled using a 35 K single-stage cryocooler. 

Light (white (10 mW), red ($\lambda=633$ nm, 1.4 mW), green ($\lambda=532$ nm, 1.4 mW) or blue ($\lambda=405$ nm, 1.4 mW)) was sent through a multimode optical fiber to illuminate the surface of the semiconductor sample as shown in Fig. 1. The microwave response in the sample was characterized by a change in the measured frequency and losses of a suitably chosen whispering gallery mode that has a sufficiently high order azimuthal mode number such that its frequency and microwave losses are largely determined only by the semiconductor and not the enclosing cavity. The resonance was weakly coupled to a vector network analyzer and tracked with a fast data acquisition system. 

\begin{figure}
\includegraphics[width = 3.5 in]{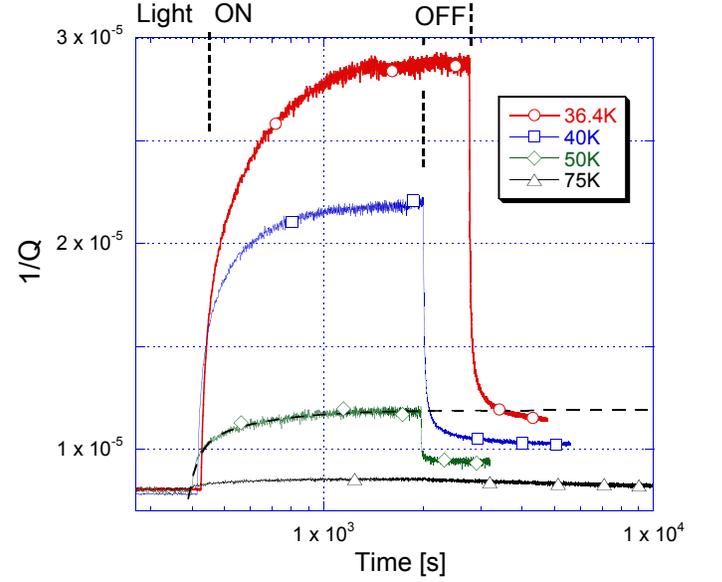}
\caption{\label{fig:fig3} (Color online) Time evolution of the microwave losses ($Q^{-1}$) of GaP illuminated with a 10 mW white light source at different temperatures. Each data set has been scaled by a constant offset to make the initial $Q^{-1}$ equal to the same initial value as the 50 K measurement, before illumination. The (black) dashed line is a modeled fit. See text for details. }
\end{figure}

The observed change of the mode frequency ($\Delta f$) is related to a modification of the permittivity of the semiconductor where,
\begin{equation} \label{eqn:fshift}
\frac{\Delta f}{f_0} = - \frac{p_{\varepsilon}}{2}\frac{\Delta \varepsilon_r}{\varepsilon_r},
\end{equation}                                                    
and $\Delta f = f-f_0$, $f_0$ is the frequency of the mode before illumination, $p_{\varepsilon}$  is the electric energy filling factor,  $\varepsilon_r$ the isotropic real dielectric permittivity and  $\Delta \varepsilon_r$  the change in this value as compared to that before illumination or after the illumination was switched off, whichever the case. 

A change in the microwave losses ($tan \delta$) is related to a change of the conductivity, 
\begin{equation} \label{eqn:tand}
tan\delta = tan\delta_d + \frac{\sigma_{0}}{\omega \varepsilon_0 \varepsilon_r}+\frac{\Delta \sigma}{\omega \varepsilon_0 \varepsilon_r},
\end{equation}
where $tan \delta_d$  is the dielectric loss tangent associated with pure dielectric loss mechanisms (e.g., electronic and ionic polarization),  $\sigma_0$ is the electric conductivity in the semiconductor in darkness, and  $\Delta \sigma$  is the excess conductivity associated with the change of state of free carriers during and after illumination. The latter may be determined from, 
\begin{equation} \label{eqn:condshift}
\frac{\Delta \sigma}{\omega \varepsilon_0 \varepsilon_r} = \frac{1}{p_{\varepsilon}} \left( \frac{1}{Q}-\frac{1}{Q_0} \right),
\end{equation}  
where $Q$ is the mode Q-factor after illumination and $Q_0$ is the Q-factor of the mode before any illumination. 

Since the mode electric energy filling factor is constant over our measurements, and nearly equal to unity in a whispering gallery (WG) mode with a large azimuthal mode number, the reciprocal of the Q-factor is a good approximation for the microwave losses here.

Different  WG modes were chosen and measured in the different samples at certain cryogenic temperatures. The $WGE_{11,0,0}$ mode with a resonance frequency of 10.9 GHz at 50 K was measured in the GaP sample, the $WGH_{7,0,0}$ mode (12.53 GHz at 50 K) in the GaAs sample and the $WGE_{8,0,0}$ mode (36.6 GHz at 50 K) in the 4H-SiC sample.

The resonators were cooled, the temperature stabilized at a fixed set point, the light source was turned on and remained until a stable state (both in mode frequency and losses) was obtained then the light source was switched off. The frequency and Q-factor of the chosen WG mode were tracked by the data acquisition system during this process. 

Figures 2 and 3, respectively, show the time evolution of the relative permittivity and losses of GaP at different temperatures as a function of time from when the white light source is turned on and then after it is turned off. After the light source was turned on we observed an increase of both permittivity ($\varepsilon_r$) and microwave losses ($Q^{-1}$).  After a few minutes, a stable state is reached, then the light source was turned off and the measured parameters tended to return towards their initial values. Because of a persistence of photoconductivity \cite{Dissanayake2, Dissanayake3, Jeanjean, Brunthaler, Dang, Hamed, Hartnett1, Hartnett2} in the sample, the final state is different from that of the initial dark condition. And additionally it is worth noting that at lower temperatures the measured changes are larger. The effect is discussed in detail later.

The time evolution of the microwave losses in GaP and the relative permittivity in 4H-SiC illuminated with various monochromatic sources (red, green and blue) at 50 K, and the same light power, are shown in Figs 4 and 5, respectively. Finally, in Fig. 6 we show the time evolution of the microwave losses in GaAs under red and blue light at 50 K. 

\section{Proposed phenomenological model}
The change in the measured permittivity, resulting from a change of polarization state of the semiconductor, due to trapping or detrapping of free carriers, can be modeled\cite{Hartnett1} as a function of time $t$, from some initial time $t_0$ before switching, as,
\begin{equation} \label{eqn:epsfit}
\frac{\Delta \varepsilon_r}{\varepsilon_r} = a_{\infty} + \sum_{i=1}^n b_i \left (\mbox{exp}[(t-t_0)/\tau_i]-1 \right)^{-1},
\end{equation}
where $a_{\infty}$ is the steady-state limiting value of the change in the permittivity after a long time period with the light source ON or OFF, $n$ a small positive integer allowing for each trapping process, $\tau_i$ are characteristic time constants, which along with $b_i$ and $t_0$  are free parameters determined from the best fit to the data. When $b_i$ is positive the term applies to a retrapping process and when it is negative to a detrapping process in Eq. (\ref{eqn:epsfit}).  

Similarly the change in the measured microwave losses,  resulting from a change in the conductivity of the semiconductor, due to trapping or detrapping of free carriers, can be modeled as a function of time $t$, from some initial time $t_0$ before switching, as,
\begin{equation} \label{eqn:recipQfit}
\frac{1}{Q}-\frac{1}{Q_0} = l_{\infty} + \sum_{i=1}^n c_i \left (\mbox{exp}[(t-t_0)/\tau_i]-1 \right)^{-1},
\end{equation}
where $l_{\infty}$ is the steady-state limiting value of the change in the microwave losses after a long time period with the light source ON or OFF, and $c_i$ are free parameters determined from the best fit to the data. As above a positive valued $c_i$ applies to a retrapping process and when negative to a detrapping process in Eq. (\ref{eqn:recipQfit}).

When Eqs (\ref{eqn:epsfit}) and (\ref{eqn:recipQfit}) are applied to the data of Figs 2 and 3 they agree to within a small constant factor. Hence we can write
\begin{equation} \label{eqn:equal}
\frac{1}{Q}-\frac{1}{Q_0} = \frac{\alpha_{\lambda}}{2}\frac{\Delta \varepsilon_r}{\varepsilon_r},
\end{equation}
where $\alpha_{\lambda} \approx 0.4$ but depends on the incident light wavelength. Hence Eq. (\ref{eqn:equal}) holds where one sums over  red and green light contributions in GaP, for example, as shown in Fig. 4, and fit it to the 50 K data of Figs 2 and 3. Equation (\ref{eqn:equal}) holds where  the temperature and laser power is the same but since the white light source is polychromatic the coefficient $\alpha_{\lambda}$ allows for different power at different wavelengths and the microwave losses at those wavelengths. 

Our model  Eq. (\ref{eqn:recipQfit}) was applied to the red and green wavelength measurements of the microwave losses at 50 K as shown in Fig. 4. The best fit curves (with statistical residuals $R > 0.999$) involve two detrapping processes under the red laser light and one detrapping process under the green laser light.  No effect is observed under blue light. For the model to fit to the `Red' curve in Fig. 4 a fast and a slow detrapping process both had to be included. See Table I where the time constants are listed. 

Using these time constants for GaP we fit Eq. (\ref{eqn:epsfit}) to the 50 K permittivity measurements of the same mode under white light as shown by the middle dashed curve in Fig. 2 (labeled 50K). The $b_i$ coefficients ($i = 1,2,3$) from Eq. (\ref{eqn:epsfit}) were determined as free parameters. Therefore the fitted equation involved three terms, one for each detrapping process, two under red and one under green light. Then Eq. (\ref{eqn:recipQfit}) was fitted to the 50 K loss measurements of the mode under white light as shown by the dashed curve in Fig. 3. In this case the same $b_i$ coefficients were used but we introduced two different scaling factors $\alpha_{\lambda}$, as free parameters, where $$c_1 = \frac{\alpha_{\lambda (R)}}{2}  b_1, \, c_2 = \frac{\alpha_{\lambda (R)}}{2}  b_2, \, c_3 = \frac{\alpha_{\lambda (G)}}{2}  b_3,$$ for each of the three detrapping terms in Eq. (\ref{eqn:recipQfit}). This meant the two $\alpha_{\lambda}$'s (indicated by (R) and (G)) were the only free parameters, which yielded $\alpha_{\lambda (G)} \approx 0.36$ under green light and $\alpha_{\lambda (R)} \approx 0.48$ under red light. There is no contribution from any blue light detrapping process.

The fact that Eq. (\ref{eqn:equal}) holds at different temperatures tells us  that the same carriers that contribute to the change in conductivity in the sample produce a change in the polarization state of the sample as they leave those traps, hence ionizing the bulk.

As a result we may examine  the response of the sample to light from either its microwave losses or its permittivity to determine the physics involved. In some cases there was better resolution in the frequency measurements than in the Q-factor measurements and hence in this paper we have used the better data where appropriate.

\begin{figure}
\includegraphics[width = 3.5 in]{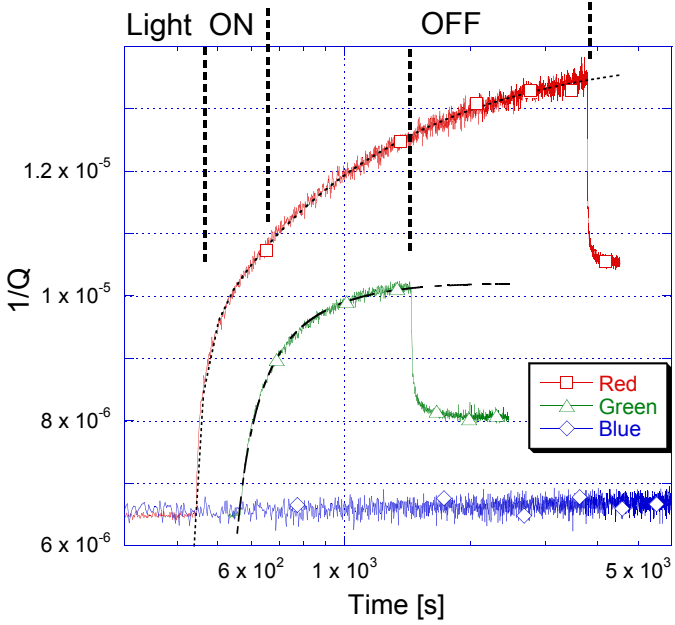}
\caption{\label{fig:fig4} (Color online) Time evolution of the microwave losses ($Q^{-1}$) of GaP at 50 K illuminated with red, green and blue light at 50 K. The (black)  dashed lines are curve fits. See text for details.}
\end{figure}
\begin{figure}
\includegraphics[width = 3.5 in]{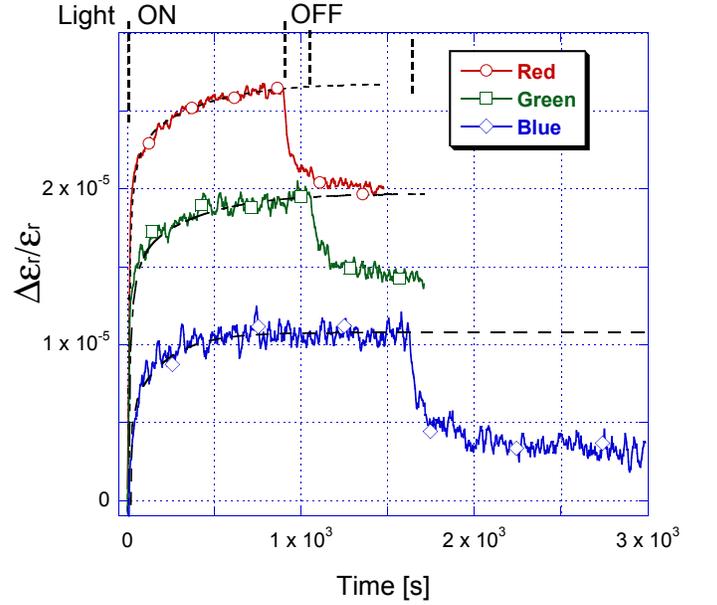}
\caption{\label{fig:fig5} (Color online) Time evolution of the relative permittivity ($\varepsilon_r$) of 4H-SiC from its initial value when illuminated with red, green and blue light at 50 K. For clarity of display the ``Green'' curve has been shifted down by $3.2 \times 10^{-6}$ and the ``Blue'' curve by $1.2 \times 10^{-5}$. The (black) broken and dashed lines are curve fits. See text for details.}
\end{figure}

\begin{table}
\begin{center}
\caption{Time constants in GaP, 4H-SiC and GaAs  under 1.4 mW red, green and blue light at 50 K, after the light source is switched on and off  (from Figs 4, 5 and 6). Each time constant is listed with its statistical 1$\sigma$ standard errors. Symbol $\lambda$ indicates the color of the laser, `r' indicates retrapping, and `d' indicates detrapping.}
\begin{tabular}{cccc}
\colrule\colrule
Sample & $\lambda$ & $\tau_i$ [s] (ON) 	& $\tau_i$ [s] (OFF) \\
\hline
GaP 		& Red r		& -									& $18.08 \pm 0.66$\\
				& Red r		& -									& $366.6 \pm 17.4$\\
				& Red d		& $23.93 \pm 0.15$  & - \\
				& Red d		& $4231.2 \pm 12.9$	& - \\
				& Green r & -									& $15.48 \pm 0.54$\\
				& Green r & -									& $177.9 \pm 3.6$\\
				& Green d	& $454.9 \pm 2.0$ 	& - \\
\hline
4H-SiC 	& Red r		& $147.0 \pm 53.5$ 	& $176.3 \pm 4.6$\\
				& Red d 	& $341.4 \pm 152.5$ & -	\\
				& Green r & $147.7 \pm 47.1$ 	&	$127.0 \pm 3.8$\\
				& Green d	& $330.3 \pm 129.3$ & - \\
				& Blue r 	&  $78.3 \pm 21.7$ 	& $36.2 \pm 4.7$\\
				& Blue r 	&  - 							 	& $1026.2 \pm 49.3$\\
				& Blue d	& $216.9 \pm 71.6$ 	& - \\	
\hline	
GaAs 		& Red	r 	& $300.6 \pm 1.7$ 	& $2.07 \pm 0.0006$\\
				& Red r 	&	-									& $221.0 \pm 0.7$ \\
				& Red d 	& $21.67 \pm 0.25$ 	& - 							\\
				& Blue r 	& - 								& $3.164 \pm 0.002$ \\
				& Blue r 	& - 								& $1539.3 \pm 4.5$\\
				& Blue d 	& $1.88 \pm 0.04$ 	& - \\
				& Blue d	& $1683.8 \pm 27.5$ & - \\
\colrule
\end{tabular}
\end{center}
\end{table}

\subsection{Detrapping and retrapping relaxation times}
The model equations (\ref{eqn:epsfit}) and (\ref{eqn:recipQfit}) provide  extremely good fits to all the measured data. In Table I lists the relaxation time constants in GaP, GaAs and 4H-SiC under red, green and blue light, after the light source is switched ON and OFF (the table is labelled accordingly), where it was possible to obtain data. Equation (\ref{eqn:recipQfit})  was fitted to data in Figs 4 and 6 to determine those time constants for the GaP and the GaAs samples. Equation (\ref{eqn:epsfit}) was fitted to data in Figs 5 to determine those time constants for the 4H-SiC sample. The curve fits were applied to data both where the light sources were turned on and after they were turned off. Not all fits are shown in the figures. 

In GaP we saw two detrapping processes (under red light) after the light was turned on, and also two retrapping processes (under both green and red light) after the light was turned off. This is indicated by the symbols `r' and `d' respectively in Table I. To use only one detrapping or retrapping term to achieve a fit to the data in these cases was very poor. Similarly this was required also with GaAs under blue light illumination. In the 4H-SiC and GaAs samples we saw retrapping occurring along with detrapping.   Most significantly in GaAs under red light illumination we see a fast detrapping process accompanied by a slower but significant retrapping process.

\begin{figure}
\includegraphics[width = 3.5 in]{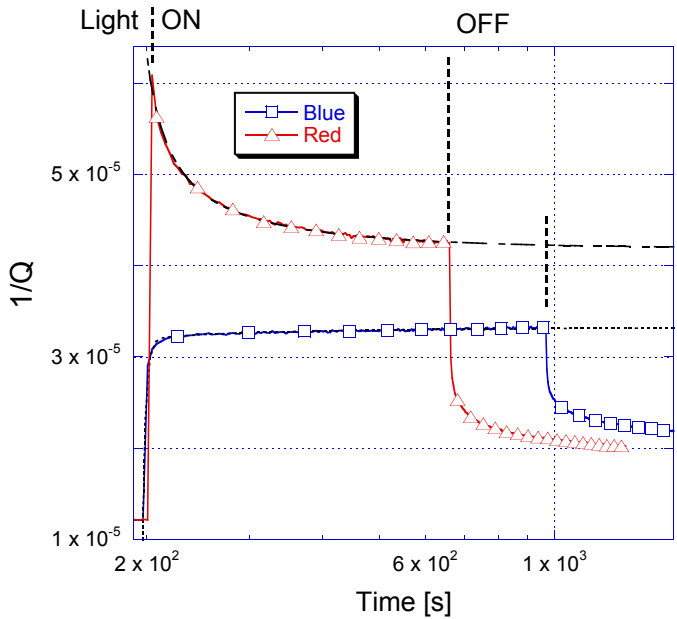}
\caption{\label{fig:fig6} (Color online) Time evolution of the microwave losses ($Q^{-1}$) of GaAs illuminated with red and blue light at 50 K. The (black) dotted and dashed lines are curve fits. See text for details. }
\end{figure}

\begin{table}
\begin{center}
\caption{Fit coefficients $b_i$ ($i = 1, 2$) in Eq. (\ref{eqn:epsfit}) for GaP sample illuminated with a 10 mW white light source at different temperatures, determined from curve fits in Fig. 2. R is the statistical residual for the fit in each case.}
\begin{tabular}{ccccc}
\colrule\colrule
Temp. [K] 	& $b_1$	& $b_2$ & R \\
\hline
36.4				& $9.3 \times 10^{-2}$ 	& $2.2 \times 10^{-5}$	& 0.9979\\
50 					& $2.8 \times 10^{-3}$	& $3.7 \times 10^{-6}$	& 0.9978\\
75 					&	$2.1 \times 10^{-5}$	& $8.1 \times 10^{-7}$	& 0.9980\\
\colrule
\end{tabular}
\end{center}
\end{table}

\subsection{Temperature dependence}
To explain the effects observed as a function of temperature, we hypothesize that proportionally more carriers become trapped on impurity/defect sites at lower temperatures. Therefore we see a reduction in microwave losses as the intrinsic conductivity in the semiconductor is reduced. The initial `in darkness' $tan\delta$ reduces with temperature. \cite{Krupka}  When the light source is turned on, the energy given by the incident photons produces a change of state of the impurity/defect sites with energies in the forbidden band gap generating the free carriers. At room temperature, the thermal energy is so high that all these sites with energies in the band gap are ionized.

This is demonstrated by the results in Figs 2 and 3. (Note: $Q^{-1}$ in Fig. 3 has been scaled by a constant offset to represent all data sets starting from the same initial value.) We took Eq. (\ref{eqn:epsfit}) with 3 fixed detrapping time constants for the GaP sample at 50 K (determined from fits in Fig. 4), $\tau_1 = 23.9$ s, $\tau_2 = 454.9$ s  and $\tau_3 = 4231.2$ s and also fixed $b_3 = 4 \times 10^{-8}$ for the slowest detrapping process, and with essentially only two free parameters ($b_1$, and $b_2$), fitted to the 75 K and 36.4 K data of Fig. 2, after the light source is switched ON. The best fit coefficients are listed in Table II. An increase of the $b_i$ coefficient means an increase in the density of traps.

For both detrapping processes (one dominant under red and the other under green light) the sensitivity for detrapping is enhanced as the sample is cooled. Simply put,  as the sample temperature is raised more free carriers are released from traps by thermal energy, which results in a lower density of trapped carriers, thus  a reduced sensitivity to the incident photons.

\section{Energy levels}

\subsection{GaP}
The energy of the band gap in GaP is $E_g = 2.33$ eV at 50 K. \cite{Panish} According to a band-to-band transition of the free carriers (from the valence band to the conduction band or vice versa), the energy of an incident photon ($E_{ph}=h c / \lambda$) must be higher than the energy of the band gap, i.e., $E_{ph} > E_g.$ The red, green and blue photons, respectively, have energies of $1.92$ eV, $2.33$ eV  and $3.07$ eV, which are lower, equal to and higher than the band gap of the GaP, at 50 K. 
 
In Fig. 4, we observe that the blue light does not produce any change in the losses in GaP at 50 K but the red wavelengths produced the biggest increase. This suggests that the band-to-band transition is not responsible for the change of the microwave properties in the GaP sample as the energy of a red photon is lower than the band gap in GaP. Therefore, most likely unidentified impurities and defect sites with energies in the band gap must be sensitive to red photons, which detrap free carriers.

The model suggested here describes two detrapping processes for red photons -- an initial fast detrapping process into a trap which has a lower density than a slow background detrapping process.  In the case of green photons a different impurity and/or defect site is responsible for its intermediate detrapping rate. 

\subsection{4H-SiC}
Figure 5 shows the evolution of the relative permittivity of 4H-SiC under red, green and blue light at 50 K. The change in relative permittivity under illumination seems to be almost independent of wavelength. All three data sets in Fig. 5 lie close to each other, however for clarity of display `Green' and `Blue' curves have been shifted down. 

The energy of the band gap in 4H-SiC is $E_g = 3.26$ eV at 50 K. \cite{Persson} So the energy of all incident photons is  too small for a band-to-band transition in the semiconductor. In this case, the similar response to different wavelengths can be explained by multiple low density  impurity/defect trapping sites  with ionization energies  equivalent  to the different wavelengths. The detrapping and retrapping time constants under red and green laser light are almost identical but under blue light there are faster detrapping and retrapping processes.  This suggests it is the same impurity/defect which responds to green and red light but a quite different species that responds to the blue.

The model suggested here describes two trapping processes -- a slow background detrapping process, with an initial fast retrapping process into a trap which has a lower density than the former. 

\subsection{GaAs}
Figure 6 shows the evolution of the losses in GaAs under red and blue wavelengths at 50 K. The sample was also measured under green light but, due to the very high sensitivity of the WG mode at these wavelengths it was not possible to record the evolution of the resonance mode. The mode rapidly was lost from the measurement system with a time constant of the order of a few milliseconds. 

The energy of the band gap in GaAs at 50 K is $E_g = 1.51$ eV. \cite{Panish} In this case, the photons incident on the sample all have energies higher than the band gap of the semiconductor, hence it could be that these photons excite electrons from the valence band to the conduction band. However, the sample is ultra sensitive to illumination with the green light, yet much less so to blue light at higher photon energy. This suggests that the response is due, at least in part, to a much larger density of impurity/defect sites sensitive to green light as compared to those sensitive to red and blue wavelengths. Moreover, we observe a difference in the change of excess conductivity when the sample is measured under red and blue light. Note the initial rapid rise in the microwave losses ($Q^{-1}$) under red light. By comparison, in both GaP and 4H-SiC (Figs 4 and 5) we observed a slow increase in the relative permittivity or the microwave losses of the material. In GaAs we observed a sharp jump in the excess conductivity followed by a slow decrease to a persistent value. 

The model suggested here describes two trapping processes -- an initial, fast, detrapping process with a slower retrapping process, into a trap which has a lower density than the former, so that it eventually saturates without all the carriers being retrapped. When the light source  is repeatedly switched on and off the density of carrier involved in the fast detrapping process is reduced against a background of carriers trapped on impurity/defect sites with the slower retrapping process.

In general the responses seen in all materials are strongest under either red or green light. In part, this may be due to the penetration depth that these wavelength achieve in the semiconductor material. Both GaP and GaAs are dark grey colored but 4H-SiC is almost transparent. The latter is consistent with it passing all wavelengths, hence, though low optical absorption,  similar responses in Fig. 5. In GaP though the blue light is strongly absorbed, few impurity/defect sites see the radiation, which is consistent with no response to blue light in Fig. 4. 

\section{Concluding remarks}
We have proposed a phenomenological model that suggests that the impurity/defect sites with ionization energies in the band gap of the measured semiconductors are consistent with the observed changes of their microwave properties (permittivity and losses) under illumination at cryogenic temperatures. At room temperature, nearly all these traps are thermally ionized. As the sample temperature is lowered past 100 K free carriers are trapped on those sites. Hence the lower the temperature the more sensitive to photo-ionization do those traps become. Their sensitivity is then dependent on their respective ionization energies and their density within the semiconductor.
\\
\section{Acknowledgment}
The authors wish to thank Gia Parish and Bob Stamps for their helpful suggestions and comments. This work was supported by the Australian Research Council.

\end{document}